\documentstyle{article}
\begin{document}
\large
\baselineskip=16pt

\vsize=17. true cm
\hsize=13.5 true cm
\hoffset=0.9 true cm
\tolerance=10000
\hfuzz=10pt

\title{Conservation Laws in Field Dynamics or Why Boundary Motion is Exactly
Integrable?}
\author{Mark B. Mineev--Weinstein\\[.3in]
Center for Nonlinear Studies, MS--B258\\Los Alamos National Laboratory\\
Los Alamos, NM 87545\\[.3in]
tel: 1(505)667-9186; fax: 1(505)665-3003;\\ e-mail: mineev@goshawk.lanl.gov}

\maketitle
\begin{abstract}
An infinite number of conserved quantities in the field dynamics $\phi_t =
L\,U(\phi) + \rho$ for a linear Hermitian (or anti-Hermitian) operator $L$, an
arbitrary function $U$ and a given source $\rho$ are presented. These integrals
of motion are the multipole moments of the potential created by $\phi$ in the
far-field. In the singular limit of a bistable scalar field $\phi = \phi_{\pm}$
(i.e. Ising limit) this theory describes a dissipative boundary motion (such as
Stefan or Saffman-Taylor problem that is the continuous limit of the
DLA-fractal
growth) and can be exactly integrable. These conserved quantities are the
polynomial conservation laws attributed to the integrability. The criterion for
integrability is the uniqueness of the inverse potential problem's solution.

\end{abstract}
\vspace{.3in}
\hspace{.4in}
\pagebreak

{\it Motivation:}
Two-dimensional (2D) Laplacian growth (known also as Hele-Shaw, Saffman-Taylor,
or the quasistatic Stefan problem) has many attributes of an exactly
integrable nonlinear problem [1]. At the same time it is dissipative, which is
not usually the case for exactly integrable theories. These problems of moving
boundaries belong to the more general class of pattern formation phenomena
which have been studied intensively for the last decade [2]. From different
first-principles models describing pattern formation we will choose and discuss
here the so-called phase-field model [3] derived from the basics of the
thermodynamics and describing the kinetics of various phase transitions, such
as different kinds of solidification (from the supercooled liquid, from the
supersaturated solution, from the electrolyte), phase transformations in liquid
crystals and in other condensed matter systems, etc. It is known that this
phase-field model in the appropriate singular limit is equivalent to the
symmetrical Stefan problem [4]. The following elementary (but unexpected)
observation motivated this work: a quite general field dynamics (Eq.(1) below)
appears to have an infinite number of conservation laws such that in the
singular limit the appropriate interface dynamics can be an exactly integrable
(dissipative) theory. Besides its purely fundamental interest, this result can
help one to study of cases with nonzero (and even anisotropic) surface tension.
It turns out that the conservation laws in the presence of surface tension are
not lost, but are only corrected (Eq.(35) below). Thus in the framework of
exact
integrability one can eliminate the short-wave instability, which is of great
physical and practical importance, because without this stabilization the
problem of Laplacian growth is ill-posed [2].

{\it Conserved integrals:}
Let us consider the following dynamics of a field $\phi(x,t)$ on the
$D$-dimensional manifold $M$ $(x \epsilon M)$:
$$\phi_t = L\,U (\phi) + \rho\,\eqno(1)$$
where $\phi_t$ means partial derivative of the $\phi$ with respect to time $t$,
$L$ is a linear Hermitian or anti-Hermitian operator, $U$ is an arbitrary
(generally nonlinear) function of $\phi$ (and may also be a function of
$\phi_t$
and of $\nabla\phi$), and $\rho(x,t) $ is a given source term.

Let us consider solutions, $\psi_n (x)$, of the equation
$$L\,\psi_n = 0\,\eqno(2)$$
such that the integrals
$$c_n(t) = \int_M\psi_n\,\phi\,d^Dx\,\eqno(3)$$
are finite.

Here if $D>1$, the zero eigenvalue of $L$ is infinitely degenerate, where  $n$
can be discrete or continuous. In general, $n$ stands for the appropriate set
of
numbers designating the eigenfunctions of zero-eigenvalue of $L$. For example,
if $L$ is the Laplacian and $D=2$ we have

$$\psi_n = z^{\pm n}\,\eqno(4)$$

Here $n=0,1,2,...,\infty$;~~and $z$ is a complex coordinate on the plane
$z=x+iy$. For $D=3$,
$$\psi_n^{(m)} =  P_{n-1}^{(m)}(\theta)\,e^{im\omega}/r^n\,\eqno(5)$$
or $$=\,P_n^{(m)} (\theta)\,e^{im\omega}\,r^n\,\eqno(6)$$
Here, $r,\,\theta$, and $\omega$ are the polar coordinates and
$P_l^{(n)}(\theta)$ are the associated Legendre polynomials with $|m|<n$.
Also $n>0$ in Eq.(5).

One can immediately see that the dynamics of the integrals $c_n$ is known and
given by
$$dc_n/dt = \int_M{\psi_n\,\phi_t\,d^Dx}$$

$$= \int_M{\psi_n\,L\,U (\phi)\,d^Dx} + \int_M{\psi_n\,\rho\,d^Dx}
 = \int_M{\psi_n\,\rho\,d^Dx}\,\,\,\,.\eqno(7)$$
The first integral vanishes because $L$ is Hermitian (or anti-Hermitian):

$$\int_M{\psi_n\,L\,U (\phi)}d^Dx = \pm \int_M{U (\phi)L\,\psi_n}\,d^Dx = 0
\,\,\,\,.\eqno(8)$$

There are two important particular cases:

A) $\rho$ is constant in time. In this case the integrals $c_n$'s are linear in
time.

B) $\rho = 0$. In this case the $c_n$'s are constants in time.

{\it Hamiltonian and dissipative processes:}
The dynamics (1) is quite general. Many important physical processes are
described by this equation with $U = \delta H/\delta\phi$, where (in the
language of the theory of phase transitions) $H[\phi]$, $\phi$, and $U$ are the
free energy functional, the order parameter, and the external field
respectively. We do not specify the nature of $\phi$ yet: it can be a scalar,
vector, tensor, etc. For the $\rho=0$ case, the corresponding dynamical
equation is:
$$\phi_t = L\,\,{\delta H\over \delta\phi}\,\eqno(9)$$
and can describe both Hamiltonian and dissipative processes. In the
former case $H$ is a Hamiltonian, in the latter one $H$ can be a Lyapunov
functional, i.e. a functional decreasing in time.

An example of a Hamiltonian process is:
$$\phi_t=\nabla \,\times\,{\delta H\over \delta\phi}\,\eqno(10)$$
Here $\phi$ is a vector and $L$ = {\bf curl}. In this case the zero eigenvalue
functions are  $\psi$ = {\bf grad}\,\,$W$ with arbitrary $W$, and it is easy to
see that $H$ is constant in time:
$$ {dH\over dt} = \int {\delta H\over \delta\phi}\,\,\,\phi_t\,\,\,d^Dx =\int
{\delta H\over \delta\phi}\,\,\,[\nabla\,\times\,{\delta H\over
\delta\phi}]\,\,\,d^Dx = 0\,\eqno(11)$$

An example of a dissipative process is:
$$\phi_t = (a\nabla^2-b)\,{\delta H\over \delta\phi}\,\eqno(12)$$
Here $\phi$ is a scalar and $L=a\nabla^2-b$, where $a$ and $b$ are positive
real numbers. It is easy to see that if there is no source ($\rho=0$), $H$
decreases in time:
$$ {dH\over dt} = \int {\delta H\over \delta\phi}\,\,\, \phi_t\,\,\,d^Dx =\int
{\delta H\over  \delta\phi}\,(a\nabla^2-b)\,{\delta H\over \delta\phi}
\,\,\,d^Dx$$
$$=-\int\{a\,(\nabla\,\,{\delta H\over \delta\phi})^2+b\,({\delta H\over
\delta\phi})^2\}\,\,\,d^Dx < 0\,\eqno(13)$$

{\it Mathematical interpretation of the quantities $c_n$'s}: For the particular
case where Eq.(1) describes a Hamiltonian process, $L$ is skew-symmetric, and
a Poisson bracket $\{\,,\,\}$ can be introduced. The integrals $c_n$'s are
Casimirs [5], i.e. such functionals, that commute with all others:
$\{c_n,F\}=0$ for all $F$. Since we are mainly concerned with {\it dissipative
processes and $L$ in this case, as a rule, is not skew-symmetric}, we do not
have such an
interpretation of the $c_n$'s. We can only say that the set of $c_n$'s
determines the projection of the field $\phi$ onto the null space of the
operator $L$, (i.e. the space spanned by the
eigenfunctions $\psi_n$ of $L$ with zero eigenvalue). The knowledge of the
$c_n$'s provides only partial information about $\phi$. In the source-free
case, the dynamics of $\phi$ occurs outside the null space while the projection
of $\phi$ onto the null-space of $L$ is constant in time. So a great deal of
information concerning the field $\phi$ is beyond our knowledge if the $c_n$'s
are all that we know.

  {\it Physical interpretation of the $c_n$'s}: If $L$ is a second order
elliptic operator, the quantities $c_n$'s are the coefficients of the multipole
expansion of the potential $\Phi(t,x)$ created by a distribution of ``matter''
with density $\phi^{\prime}=\phi-\lim_{|x|\to\infty}{\phi}$ in the far-field:
$$L\,\Phi=\phi^{\prime}\eqno(14)$$
To see this, we write the solution of Eq.(14) as
$$\Phi(x)=\int G(x,y)\,\phi^{\prime}(y)\,d^Dy\,\eqno(15)$$
where $$L\,G(x,y)=\delta(x-y)\,\eqno(16)$$
and expand the Green's function $G(x,y)$ as
$$G(x,y)=\sum_n^{ }\psi^{'}_n(x)\psi_n(y)\,\eqno(17)$$
where $\psi^{'}_n$ and $\psi_n$ are the two linearly independent solutions of
the equation $L\,\psi=0$ corresponding to a given $n$. Assuming the
commutativity of the sum and the integral, we finally get
$$\Phi(t,x)= \sum_n^{ } c_n(t)\, \psi^{'}_n(x)\,\eqno(18)$$

Let us take the Laplacian cases $L=\nabla^2$ for $R^2$ and $R^3$
and assume the field $\phi^{\prime}(x)$ to be localized so that it is zero in
the far-field. In the 2D case
$$\Phi(t,z)= Re\,\sum_n^{ } c_n(t)\, z^{-n}\,\eqno(19)$$
where
$$c_n(t)=\int_{R^2} z^n\,\phi(t;x,y)dx\,dy\,,\eqno(20)$$
and in the 3D case
$$\Phi(t,x)=\sum_n^{ } \sum_m^{ } c_{nm}(t)\, P_n^{(m)}(\theta)\,e^{im\omega}
/r^{(n +1)}\,\eqno(21)$$
here $x$ stands for a vector-position and
$$c_{nm}(t)=\int_{R^3}\,P_n^{(m)}(\theta)\,e^{im\omega}r^n\,\phi(t; r, \theta,
\omega)\, d^3x\,\eqno(22)$$

Now the question arises: how much can we say about the density of ``matter''
$\phi(t,x)$ from the knowledge of the potential created by $\phi$ in the far
field? This is a classical inverse potential problem having in the general
case an infinity of solutions. One can imagine for example a charged ball with
initially a spherically symmetrical but inhomogeneous charge density
$\phi(0,r)$.
It is clear that all the $c_n$'s except $c_0$ are zero in this case while
$\phi(t,r)$ can be distributed arbitrarily. So we have an infinite number of
solutions for $\phi$ satisfying the infinite number of constraints $c_n=c_0\,
\delta_{0,n}$ where $\delta_{0,n}$ is the Kronecker-symbol, and so the same
constants $c_n$'s can correspond to different fields $\phi(t,x)$ (see ref.[6]).

{\it Stefan problem as a singular limit of the ``asymmetric'' phase-field
dynamics}:
 Now we consider the process described by the following equations:
$$\cases{\phi_t + \beta\,\,U_t={1\over 2}\,\nabla [((D_+ + D_-)+(D_+ -
D_-)\phi)\,\nabla U] \cr \alpha\,\xi^2\,\phi_t\,=\,\xi^2\,\nabla^2\,\phi + {
(\phi-\phi^3)/a} + U}\eqno(23)$$
This equations encompass several well-known problems. If $D_+ = D_-$ the system
(23) coincides with the phase-field model
[3] which describes a broad class of pattern formation problems where the
order parameter $\phi$ is called a ``phase''. $D_+$ and $D_-$ stand for the
diffusions coefficients in the bulk where $\phi=1$ and $\phi=-1$ respectively
(i.e. far from the ``interface'' defined as the  surface described by
$\phi(t,x)=0$). The first equation in the system (23) is a particular case of
Eq.(1) for $\beta = 0$, and the second one is the time-dependent
Ginzburg-Landau equation [7] in the presence of external field $U$. Here
$\alpha\,\xi^2$ and $\beta$ are kinetic coefficients or the
relaxation times of the fields $\phi$ and $U$ respectively, $\xi$ is the
coherence length, and $a^{-1}$ is the double-well potential depth. One can
easily see that if $\alpha=\beta=0$ and $D_+ = D_-$ the system (23) is
equivalent to the Cahn-Hilliard equation [8]
$$\phi_t=\,\,-\nabla^2(\xi^2\,\nabla^2\,\phi + (\phi -\phi^3)/a)\,\eqno(24)$$
describing the process of phase separation. It corresponds to the dynamics
described by Eq.(9) when $L=\nabla^2$ and the functional $H$ has the
Ginzburg-Landau free energy form:
$$ H\,=\,\int_M\,({\xi^2(\nabla\,\phi)^2\over 2} -\,{\phi^2\over
2a}\,+\,{\phi^4\over 4a})\,d^D\,x \eqno(25)$$

As it was shown by Caginalp [4] the phase-field model (i.e. the Eqs.(23) with
$D_+=D_-$) in the singular limit $\xi \rightarrow 0$ and $a \rightarrow 0$
reduces to the symmetrical (i.e.when $D_+=D_-$) Stefan problem
$$\beta\,U_t = \nabla^2\,U\,\,\,\,\,\,\,\,\,\,\,\,\,\,\,\,\,\,\,\,x\in M\,\,
\eqno(26a)$$
$$U=\sigma(\kappa +
\alpha\,v_n)\,\,\,\,\,\,\,\,\,\,\,\,\,\,\,\,\,\,\,\,x\in\partial\Omega(t)\,\eqno(26b)$$
$$v_n =
-[\partial_n\,U]\,\,\,\,\,\,\,\,\,\,\,\,\,\,\,\,\,\,\,\,\,\,\,\,\,\,\,x\in\partial\Omega(t)\eqno(26c)$$
where $\partial\Omega(t)$  is a moving interface between the growing bubble
$\Omega(t)$ and the rest of the space, $\sigma$ is the surface tension
proportional to $\xi/\sqrt{a}$, $\kappa$ is the mean curvature of the
moving boundary $\partial\Omega$, $v_n$ is the normal component of the local
velocity of $\partial\Omega(t)$, and $[\partial_n\,U]$ is the jump of the
normal component of the gradient of $U$, $\partial_n\,U$, across the boundary
$\partial\Omega(t)$. Laplacian growth corresponds to the quasistatic Stefan
problem (or Hele-Shaw problem) when $\beta=0$.

The advantage of the ``asymmetric'' phase-field model (23) is that the same
singular limit ($\xi \rightarrow 0$, $a \rightarrow 0$) reduces the system (23)
to the general (not necessarily symmetric) Stefan problem:

$$\beta_{\pm}\,U_t = \nabla^2 U
\,\,\,\,\,\,\,\,\,\,\,\,\,\,\,\,\,\,\,\,\,\,\,\,\,\,\,\,\,\,\,\,\,\,\,\,\,\,\,\,\,\,\,\,\,\,\,\,\,\,\,\,\,\,\,\,\,\,\,\,\,\,\,\,\,x\in\Omega_{\pm}\,\eqno(27a)$$
$$U=\sigma(\kappa +
\alpha\,v_n)\,\,\,\,\,\,\,\,\,\,\,\,\,\,\,\,\,\,\,\,\,\,\,\,\,\,\,\,\,\,\,\,\,\,\,\,\,\,\,\,\,\,\,\,\,\,\,\,\,\,\,\,\,\,\,x\in\partial\Omega\eqno(27b)$$
$$v_n = -D_+\partial_n\,U|_{phase 1}+D_-\partial_n\,U|_{phase
2}\,\,\,\,\,\,\,\,\,\,\,\,\,\,\,\,\,\,\,\,x\in\partial\Omega(t)\eqno(27c)$$
Here $\beta_{\pm}=\beta/D_{\pm}$, while $\Omega_{\pm}$ are domains
corresponding to phases $\phi={\pm}1$.

One can further extend the system (23) to the more general processes described
by equations
$$\cases{\phi_t + \beta\,\,U_t=(p\nabla q\nabla + r)\,U \cr
\alpha\,\xi^2\,\phi_t\,=\,\xi^2\,\nabla^2\,\phi + {F^{\prime}(\phi)/a} +
U}\eqno(28)$$
Here $p$, $q$, and $r$ can be functions of $\phi$ or of the space $x$. In
other words the RHS of the first equation in the (28) is the arbitrary elliptic
operator of the second order acting on $U$. In the RHS of the second equation
of (28) the function $F(\phi)$ is introduced such that its derivative with
respect to the scalar field $\phi$ has two stable solutions, $\phi_{\pm}$, and
the  unstable one, $\phi_o$: $F^{\prime}(\phi_{\pm})= F^{\prime}(\phi_0)=0$. It
is easy to see that the same singular limit as used above reduces the
system (28) to the following interface dynamics:
$$\beta_{\pm}\,U_t = (p_{\pm}\nabla q_{\pm}\nabla + r_{\pm})\,U
,\,\,\,\,\,\,\,\,\,\,\,\,\,\,\,\,\,\,\,\,\,\,\,\,\,\,\,\,\,\,\,\,\,\,\,\,\,\,\,\,x\in\Omega_{\pm}\,\,\eqno(29a)$$
$$U=\sigma(\kappa +
\alpha\,v_n)\,\,\,\,\,\,\,\,\,\,\,\,\,\,\,\,\,\,\,\,\,\,\,\,\,\,\,\,\,\,\,\,\,\,\,\,\,\,\,\,\,\,\,\,\,\,\,\,\,\,\,\,\,\,\,\,\,\,\,\,\,\,\,\,\,\,\,\,\,\,\,\,\,\,\,\,\,\,\,\,\,\,\,\,x\in\partial\Omega(t)\,\eqno(29b)$$
$$v_n =
-p\,[\,q\,\partial_n\,U]\,\,\,\,\,\,\,\,\,\,\,\,\,\,\,\,\,\,\,\,\,\,\,\,\,\,\,\,\,\,\,\,\,\,\,\,\,\,\,\,\,\,\,\,\,\,\,\,\,\,\,\,\,\,\,\,\,\,\,\,\,\,\,\,\,\,\,\,\,\,\,x\in\partial\Omega(t)\eqno(29c)$$
where $p_{\pm}$, $q_{\pm}$, and $r_{\pm}$ are the functions $p$, $q$, and $r$
respectively defined in the domains $\Omega_+$ and $\Omega_-$. It was shown [9]
that this kind of interface dynamics (describing rather ``elliptic'' than
Laplacian growth, because of $\nabla^2 \rightarrow p\nabla q\nabla + r$)
possesses the infinite number of conservation laws for an arbitrarily
dimensional process.

Now we investigate the quantities $c_n$ defined by Eq.(3) for the processes
described by (28) in this singular limit:

{\it Conserved integrals $c_n$'s in the singular limit}: The limit $\xi
\rightarrow0$ makes the width between different phases infinitesimally thin,
and the limit $a\rightarrow0$ excludes fluctuations of $\phi(x)$, so it can
take only three values: two stable, $\phi_{\pm}$, and one unstable, $\phi_0$.
Let us choose for simplicity $F$ in the Ginzburg-Landau form: $F^{\prime}=
\phi - \phi^3$. In this case $\phi_{\pm} ={\pm}1$, $\phi_0 = 0$. The equation
$\phi_0(t,x)=0$ describes the boundary moving between two phases $\phi =
{\pm}1$. By introducing the new function
$$\phi ^{\prime} = (\phi+1)/2\,\eqno(30)$$
we describe the interface $\phi(t,x)=1/2$ moving between two phases described
by
$\phi^{\prime}= 1$ in $\Omega(t)$ and $\phi^{\prime}= 0$ outside it.
Consequently, the integrals $c_n$ in this limit are:
$$c_n = \int_{R^D}\,\psi_n\,\phi^{\prime}\, dx = \int_{\Omega(t)}\, \psi_n\, dx
\,\eqno(31)$$
But these integrals  are the constants of motion found earlier in arbitrarily
dimensional boundary motion when $\alpha = \beta = \sigma = 0$ [9]. Thus, in
the singular limit $\xi\rightarrow0,\,\,a\rightarrow0$ the conserved
quantities,
$c_n$, coincide with the conservation laws previously found. Therefore the
field dynamics ((1) is a
natural extension of the boundary motion $\partial\Omega (t)$ that possesses an
infinite number of conservation laws.

{\it Integrability in the singular limit}: An infinite number of polynomial
conservation laws is a necessary (but not always sufficient) condition for a
nonlinear field problem to be exactly integrable [10]. When do the $c_n$'s
correspond to an integrable case? While the knowledge of the $c_n$'s is not
enough to recover the field $\phi(x)$, the situation is much better in the
singular limit described above where the field dynamics $\phi(t)$ is reduced to
the boundary dynamics $\partial\Omega(t)$. In this case one might recover the
interface and the problem might be solvable. Indeed here the ``matter
distribution'' $\phi^{\prime}$ is especially trivial: $\phi^{\prime} = 1$ in
$\Omega(t)$ and $\phi^{\prime} = 0$ outside it. Therefore looking for the
$\phi(t,x)$ by knowing the far-field potential $\Phi$ is reduced to the looking
for the boundary $\partial\Omega(t)$ between two phases. This problem is called
the ``inverse potential problem''[11,12] and sometimes can be explicitly
solved.
For $D=2$ and $L=\nabla^2$ using conformal mapping it is possible to recover
the
boundary $\partial\Omega$ from the knowledge of the singularities of the
function $\sum_{n=1}^{\infty} {c_n z^{-n}}$ analytic in $R^2/\Omega$ [11], and
where $c_n$ are integrals defined by (3). However, if $D\not=2$ or
$L\not=\nabla^2$ we do not have a constructive way to recover the boundary
$\partial\Omega$, but if $\Omega$ is (a) single-connected and (b) star-like
(i.e. a point inside $\Omega$ exists such that all straight rays from this
point
cross the boundary $\partial\Omega$ only once) there exists a unique domain
$\Omega$ for given potential [12]. Nevertheless it is still unclear how to
parametrize the corresponding boundary in this case. So, although the knowledge
of $c_n$'s in principle is enough to recover the boundary of a single-connected
star-like domain in the general multi-dimensional case, only the 2-D Laplacian
case is integrable in a constructive way (and not even for all kinds of
singularities of the function $\sum_{n=1}^{\infty} {c_n z^{-n}}$).

{\it Integrals $c_n$'s in the case of non-zero surface tension $\sigma=\xi/
\sqrt{a}$ and non-zero kinetic coefficient $\alpha$:} If in the singular limit
$\xi\rightarrow0 ,\,\,a\rightarrow0$ the ratio $\xi/\sqrt{a}$  has a finite
non-zero value that corresponds to a non-zero surface tension, the integrals
$c_n$ defined as in Eq.(3) generally speaking are not conserved for the
manifold $M=R^D$.(They are even not necessarily finite.) For the process
described by the system(23) with $\beta=0$ the rate of $c_m$ defined by (3) is
$$d\,c_m/d\,t = D_-\int_{S^{D-1}}\,(\psi_m\,\partial_n\,U - U\,
\partial_n\,\psi_m)\,dS^{D-1}\,\eqno(32)$$
where $S^{D-1}$ is the hypersphere in $R^D$ whose radius tends to infinity when
$\psi_m$ diverges at infinity and tends to zero when $\psi_m$ diverges at the
origin. (Without surface tension it corresponds to the ``internal'' and
``external'' one-side Stefan problems respectively, i.e. when $U$ governed by
the equation $L\,U=0$ is non-trivial only inside or outside the boundary
$\partial\Omega$, while in the complementary phase $U$ vanishes). If
$L=\nabla^2$ the function $\psi_m$ diverges (at infinity or at the origin) as
an $m^{th}$ power of $r$ (see Eqs.(4)-(6)). In this case, for the RHS of
Eq.(32)
to vanish, $U$ must decay where $\psi_m$ diverges (i.e. at infinity or at the
origin) faster than any power of $r$. Without surface tension, the field $U$
which is zero in the phase where $\psi_m$ diverges, since $U$ is a harmonic
function zero at the boundary, $U=0$. Thus in this case the RHS of Eq.(32) is
zero, so that the $c_n$'s are conserved.

For non-zero surface tension this is not the case, because $U$, satisfying
non-trivial boundary conditions (see Eq.(27b)) decays at infinity (or at the
origin) only algebraically, instead of exponentially (generally speaking as $r$
at the origin or as $r^{-1}$ at the infinity). Therefore, in this case, the RHS
of Eq.(32) is not zero, so the $c_n$'s are not conserved. For the 2D Laplacian
internal problem (when $\psi_n = z^n$) it is
$$d\,c_n/d\,t = -4D_-{\pi}nU_n\,\eqno(33)$$
where $$U(r,\theta)=\sum_{n=1}^{\infty}(r^{-n}(U_n\exp(-in\theta) +
U_{-n}\exp(in\theta))) \,\eqno(34)$$
[13]. By the same reasoning the $c_n$'s are not conserved when the kinetic
coefficient $\alpha$ is not zero.

However, there are situations when $d\,c_n/d\,t = 0$ even for non-zero surface
tension and kinetic coefficient $\alpha$: (i) First, this is the case if
instead of $L=\nabla^2$ the elliptic operator $L$ is such that its
solutions decay exponentially at infinity (for example if $L =\nabla^2 - a$
with a positive $a$). Evidently the surface integral in the RHS of Eq.(31) also
vanishes. (ii) Second, $c_n$'s are conserved for the {\it one-side } (i.e. when
$D_-=0$) non-zero surface tension case (which is beyond the phase-field model
dealing only with the symmetrical two-side model, when $D_+=D_-$). The one-side
problem with non-zero surface tension is very common for different physical
systems, for example for bubbling in Hele-Shaw cell, in solidification when one
can neglect diffusion in a solid, etc. [2].
(For simplicity we consider here only isotropic surface tension. For the
anisotropic case all the results hold, but $\sigma$ can no longer be pulled out
of integrals and derivatives since it is no longer a number but instead an
angle-dependent function.)

It is easy to obtain the rate of change of $c_n$ in this case directly from the
 boundary motion's formulation (29). Indeed because the integrand in (3) does
not depend on time, but only on the domain of integration, we have

$$d\,c_m/d\,t = \,1/dt\,(\int_{\partial\Omega(t+dt)}\,-\,\int_{\partial\Omega
(t)})\,\,\psi_m\,d\,\Gamma$$
(here $d\,\Gamma$ stands for the element of the moving surface
$\partial\Omega(t)$. Further, it equals
$$=\int_{\partial\Omega (t)}\,v_n\,\,\psi_m\,
d\,\Gamma\, =\,\int_{\partial\Omega (t)}\,\psi_m\,\partial_n\,U\,d\,\Gamma\,$$
because of Eq.(29c) (here $p=1$, $q_+=1$, and $q_-=0$ are chosen for the
simplicity);
$$=\,\,\int_{\partial\Omega (t)}\,\psi_m\,\partial_n\,U\,d\,\Gamma\,
-\,\int_{\partial\Omega (t)}\,(U\,-\,\sigma\,(\kappa\,+\,\alpha\,v_n))\,
\partial_n\,\psi_m\,\,d\,\Gamma\,$$
because of Eq.(29b), and finally
$$=\,-\sigma\int_{\partial\Omega}\,(\kappa
+\,\alpha\,\partial_n\,U)\,\partial_n\,
\psi_m\,d\,\Gamma\,\eqno(36)$$
since the vector field $(U\,\nabla\,\psi_m\,-\,\psi_m\,\nabla\,U)$ is
divergenceless because of Eq.(29a):
$$\nabla(U\,\nabla\,\psi_m\,-\,\psi_m\,\nabla\,U)\,=\,0$$

So this is the law of change of $c_m$ defined by (3) for a one-side Stefan
model in the presence of
surface tension and kinetics on the moving boundary. It is known [14] that
without sources of the field $U$ (i.e. when $\rho = 0$) the boundary $\partial
\Omega(t)$ usually has the tendency to take a circular shape in the long-time
asymptotics (excluding possible topological changes which are not considered
here). In this case we expect $d|c^0_m|/d\,t$ to be negative since $|c^0_m|$
indicates a deviation from a circular shape. It would be interesting to check
this statement in the near future.

{\it Conclusions:} We have shown straightforwardly that there exist an infinite
number of conserved quantities for the field dynamics described by Eq.(1), such
that in the case of a bistable scalar field $\phi$ the singular limit
corresponding to the boundary motion between two phases (Ising limit in the
kinetics of phase transitions) can be integrable (for instance if $D\,=\,2$ and
$L\,=\,\nabla^2$). The criterion of integrability is the solvability of the
associated inverse potential problem, where the conserved integrals $c_n$'s are
the coefficients of the multipole expansion of the fictitious potential created
by the matter distribution $\phi(x,t)$. In the presence of surface tension and
boundary kinetics these conservation laws are broken, but in the framework of
the one-side model, the evolution of quantities $c_n$'s has been calculated
(Eq.(35)). It means that in many cases when the integrals $c_n$'s completely
determine the moving cluster, we can follow the interface' evolution with
non-zero surface tension.

I am grateful to F.~J.~Alexander, R.~Almgren, and I.~M.~Gelfand for very
helpful discussions of this problem and to S.~Ponce-Dawson and P.~Sievert for
their comments on the paper.

\pagebreak

{\bf References}.

1. S.Richardson {\it J.Fluid Mech.} {\bf 56}, 609 (1972); B.I.Shraiman and
D.Bensimon {\it Phys.Rev.A} {\bf 30}, 2840 (1984); D.Bensimon and P.Pelce {\it
Phys.Rev.A} {\bf 33}, 4477 (1986); S.D.Howison {\it J.Fluid Mech.} {\bf 167},
439 (1986); M.B.Mineev {\it Physica D} {\bf 43}, 288 (1990); M.B.Mineev-
Weinstein and S.Ponce Dawson  {\it Phys.Rev.E.} {\bf 50}, R24 (1994); S.Ponce
Dawson and M.B.Mineev-Weinstein {\it Physica D} {\bf 73}, 373 (1994)

2. D.Bensimon, L.P.Kadanoff, S.Liang, B.I.Shraiman, and C.Tang {\it Rev.Mod.
Phys.}{\bf 58}, 977 (1986); {\it Dynamics of Curved Fronts}, ed. by P.Pelce
(Academic Press, San Diego, 1988)

3. J.S.Langer, in {\it Directions in Condensed Matter Physics}, (World
Scientific, Singapore, 1986) p.165

4. G.Caginalp {\it Phys.Rev. A} {\bf 39}, 5887 (1989)

5. B.A.Dubrovin, S.P.Novikoff, and A.T.Fomenko, {\it Modern Geometry - Methods
and Applications}, (Springer - Verlag, New York, 1992)

6. M.B.Mineev-Weinstein and F.J.Alexander {\it Conserved Moments in
Nonequilibrium Field Dynamics}, to be published in Journal of Statistical
Physics

7. P.C.Hohenberg and B.I.Halperin, {\it Rev.Mod.Phys.} {\bf 49}, 435 (1977)

8. J.W.Cahn {\it Acta Metall.} {\bf 9}, 795 (1961), J. D. Gunton, M. San Miguel
and P. S. Sahni,
in {\em Phase Transitions and Critical Phenomena}
ed. by C. Domb and J. L. Lebowitz,
Academic, New York, (1983).

9. M.B.Mineev-Weinstein {\it Phys.Rev.E} {\bf 43}, R2241 (1993)

10. M.J.Ablowitz and H.Segur, {\it Solitons and Inverse Scattering Transform}
(SIAM, Philadelphia, 1981)

11. G.~Herglotz, {\it \"Uber die analytische Fortsetzung des
Potentials ins Innere der anziehenden Massen}, (Teubner--Verlag, 1914)

12. P.S.Novikoff {\it Doklady AN SSSR} {\bf 18}, 165, (1938)

13. The result in the last paragraph and Eq.(33) were obtained together with
R.Almgren.

14. P.Constantin and M.Pugh {\it Nonlinearity} {\bf 6}, 393 (1993)

\end{document}